Article

# Raman study of a work of art fragment

**Barbara Federica Scremin** [1,*]

[1] National Nanotechnology Laboratory (NNL) - CNR Nanotech Institute, via per Arnesano km 5, I-73100 Lecce, Italy; E-Mail: barbara.scremin@unisalento.it, barbara.scremin@cnr.it

* Author to whom correspondence should be addressed; E-Mail: barbara.scremin@cnr.it;

Premise: The present work was not published on scientific journals because presently it is required to have informations about the whole work of art as provenience in location and about sampling. Such data were not available to the author. Therefore the author considers of importance the divulgation of the present study for the method that makes clear how is possible to carry on a Raman analysis and to extract the maximum of informations from few beginning data, which is doubtless fascinating. Procedures were examined step by step and can be easily followed thanks to the accompanying pictures. The only instrument used was a Raman microscope. It can be useful especially to students that just faced the subject of work of art investigation with Raman spectroscopy.

**Abstract:** The purpose of the present report was the study and identification of an unspecified sample on a work of art by means only of a microscope coupled to a Raman spectrometer. The origin of the fragment was unknown. The Raman spectra on the virgin sample were giving no results because of a deteriorated surface treatment, in spite of the evident blue color identified by microscopic visual inspection. The sample fragmentation and the preparation of a KBr pellet allowed the distribution of the painting layers of the different components on a flat substrate reducing surface effects. Selecting the areas of different color and focusing there it was possible to identify the pigments from their Raman spectra locally acquired by selective excitation. Raman spectra were assigned by comparison with published databases. It was possible to connect Carbon Black and Orange iron oxide, as documented historically, as constituents of Azurite preparatory layer - "Morellone", according to a technique generally employed to allow the use of Azurite blue pigment on frescos, consequently the identification of typology of work of art was deducted and attributed to a fresco.

## 1. Introduction

The context of this work is the field of composition identification on a work of art that should be performed before a restoration and conservation intervention. Research Institutions, Universities, museums Research Sections are interested in the scientific study of work of arts. In Italy they may perform independent work or collaborate with peripheral Organs ("Soprintendenza ai Beni Culturali"), with a regional character, of the Ministry of Cultural Heritage and Activities (the Culture Ministry of Italian Republic). Among the activities of "Soprintendenze ai beni Culturali" are the identification of work of arts, their finding investigation and successive binding command for their protection and control by means of specific permits on works of restoration. Moreover "Soprintendenze" make rules on work of arts transfers, exports and on projects of landscape interest. Therefore each conservation intervention, particularly on "bonded" historical artifacts including buildings, must respect specific rules established by "Soprintendenze" and the specific protocol applied in the intervention need their approval. The work of restoration must respect rules, according to the best scientific acquired knowledge and technological progresses. Works of arts periodically need conservation and restoration intervention especially if they are exposed in the natural environment and are not in a museum under strictly controlled climatic and illumination condition: materials degrade over time, and once identified, over their characteristic time. Each intervention should be reversible as a criterion, and documented. A conservation and restoration intervention first of all requires a scientific identification of materials and their assembling methods, because the respect of the original formulation and of its expressive value must be preserved as a cultural value. The new materials should match at the best the original ones chosen by the artist since they are part of the historical value of an artifact: they can tell us about the socio-economic context, the technical and philosophic evolution of an artist or of an era.

Scientific studies on work of arts are necessary preliminary steps of any action of restoration aiming at the reestablishment of the original status of the elements constituting the artwork. Now days it is praxis to document the steps of any restoration intervention aiming at the conservation and repair of cultural heritage items. Conservation actions are known to be necessary over time and a good documentation of prior actions are useful and sometimes crucial. Identifications on materials are routinely required in the area of conservation, especially when past actions are not documented.

The sample examined in this paper was not from a famous masterpiece, where the techniques and materials used by the artist deserves to be revealed to reconstruct an artist's historical path for example, but it is constituted instead an unknown deteriorated fragment.

This papers deals with the problem of identification and deserves some interest because it reports about one of such cases where the sample under study needs a destructive manipulation for a successful analysis. It will be shown how in a first attempt a straightforward application of the technique was unsuitable for the identification of the artefact composition since it was a real deteriorated sample.

The technique chosen for the identification is the Raman Microscopy. It uses a laser beam coupled to a microscope to obtain spatial resolution, thus selectivity to study the sample. The focused laser beam impinges the sample and excites the Raman vibrations of the components (in general pigments and other materials constituting the layer of interest) with a spatial resolution here of about 20 micron (the beam spot size). The collected scattered light from the sample carries the spectral information about the vibration and is analysed with a spectrometer. The outputs are Raman vibrational spectra that compared with published databases [1- 4] on reference materials may allow in general identifying the painting layer components, the pigments and their assembling.

The main results of the work are the identifications of employed pigments and the reconstruction of their stratigraphic assembly, which allowed the artefact identification typology. Specifically it could be identified that the fragment was from a fresco, from the use of the pigments stratigraphic sequence, historically reconstructed: the external layer was recognized as Azurite, the inner layers as orange iron oxide and carbon black which applied on a calcite substratum are known as components of "Morellone", a mixture that allows the use of Azurite on frescos [5]. In absence of Morellone layer the Azurite must be substituted with the more expensive Lapislazuli [5, 6].

## 2. Materials and Methods

## 2. Results

In this section results from the "path analysis" are reported. First of all the virgin sample was inspected with the microscope, then given the results from the spectroscopic analysis was then destructively manipulated to obtain spectra suitable for the Raman identification.

### 2.1. Microscopic visual inspection of the virgin fragment and analysis

Images of the fragment under the Raman microscope are presented in Figure 1. As can be observed in Figure 1 (a) the surface of the sample presented blue crystallites over a finer black crust, which indicates deterioration due to external agents, which may be chemical from the atmosphere, biological or even both together.

**Figure 1.** (**a**) Surface image of the sample. (**b**) Cross section acquired with a tilted sample position.

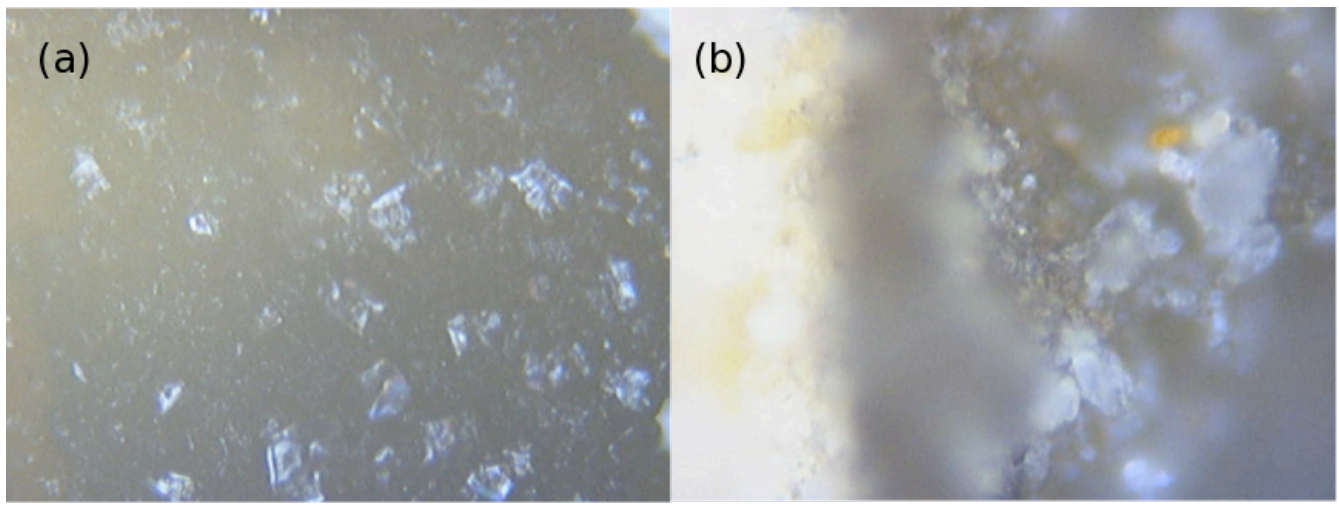

In Figure 1 (b) the fragment was tilted and the image, in spite out of focus regions appeared in a cross sectional way: starting from the left, a white-yellowish deep inner layer, a out of focus dark zone then the surface painted layer where bigger blue crystals of pigment were dispersed over a finer dark crust. It is possible to notice an orange spot over the blue crystallites. While it was possible to spectroscopically analyze the surface, the interesting cross section, due to instability of sample position was just taken into consideration for reference, since its image told about the layers different qualitative composition. In Figure 2 the sample was inspected and spectra were acquired in correspondence of the laser spot (Figure 2 a), which was translated to select different zones: the black finer crusts and the blue bigger crystallites. It is possible to notice that a 20-micron laser spot size allowed picking different portions of the surface quite selectively.

**Figure 2.** (**a**) Surface image of the sample with the laser spot for the Raman analysis. (**b**) Raman spectra acquired on different portions of the surface in (a).

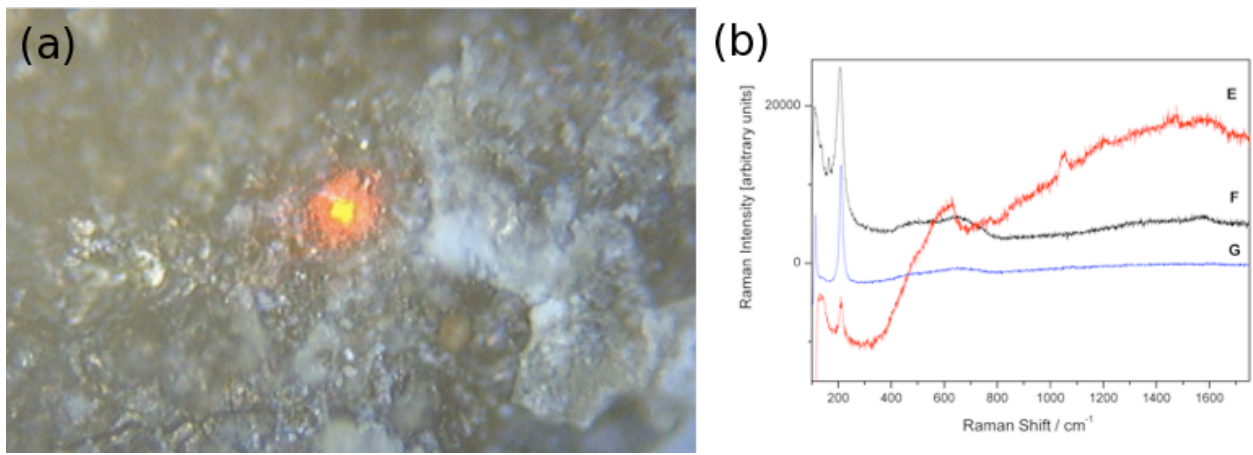

In Figure 2 (b) it is possible to observe three different Raman spectra, labeled E, F, G. The E spectrum presented a couple of bands in the low wavenumber region and few broad bands superimposed on a curved luminescence background, the latter gives indication a possible surface treatment. In F and G spectra the same two bands can be noticed: it is a too poor result to proceed to a Raman identification, in spite qualitatively evident blue bigger crystallites over the back dark crust were selectively impinged by the laser. This indicates that the surface was not virgin but deteriorated and possibly treated, for example by varnishes or oils from a previous conservative intervention, in an attempt to protect it from deterioration [5, 6].

*2.2. Resin embedded cross section and analysis*

In the field of cultural heritage research practice it is quite common to embed fragments in a resin matrix to obtain cross sections of the different painted layers. Here in Figure 3 (a) it is shown an example of this sampling technique: it is possible to distinguish a channel into which the sample is confined. The laser beam, about 20 microns in diameter was scanned along the channel in an attempt to collect data about the different layers. Comparing this image with the one of Figure 1 (b) it is imaginable to understand easily that the channel was too thin (like a 20 micron thick horizontal slice of Figure 1 (b)) for a visual inspection and for the collection of selective spectra. Anyway a trial is shown in Figure 3 (b), where the obtained spectra are reported: H spectrum is relative to the polyester matrix [7], I, L, M, N where acquired along the channel and essentially present a fluorescent background with weak signals from the matrix, therefore no information for a performable Raman identification was present also in this situation.

> **Figure 3.** (**a**) Surface image of the sample with the laser spot for the Raman analysis. (**b**) Raman spectra acquired on different portions of the surface in (a).

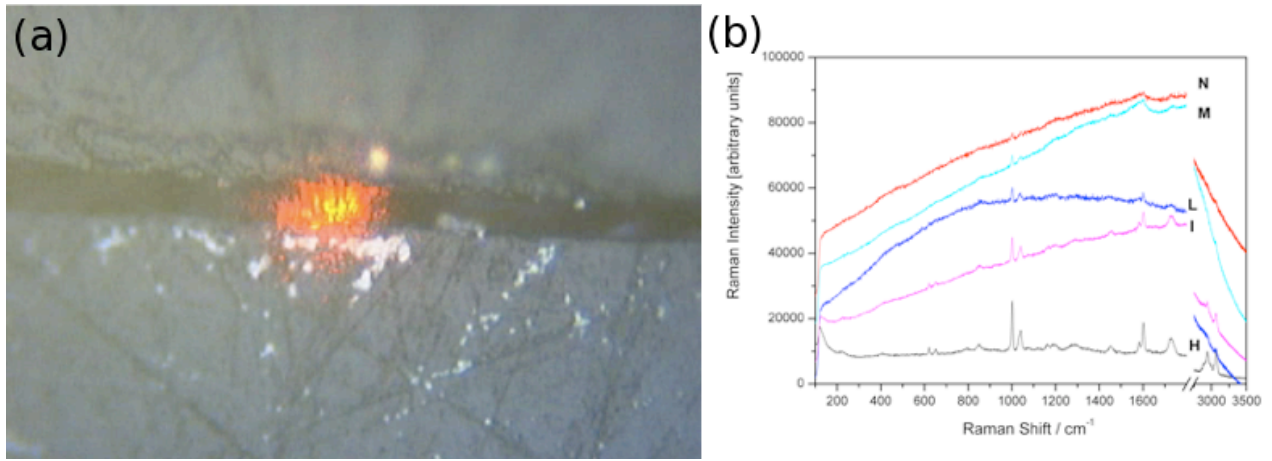

*2.3. KBr pellet technique and analysis*

Given that the above-described sampling techniques gave no useful results the sample was destructively manipulated being crushed into a mortar and mixed with KBr then pressed to obtain a pellet, shown in Figure 4 (a).

> **Figure 4.** (**a**) Image of the fragmented sample included in a KBr pellet with the highlighted region on which the Raman collection where performed. (**b**) Raman spectra acquired on different portions indicated in (a).

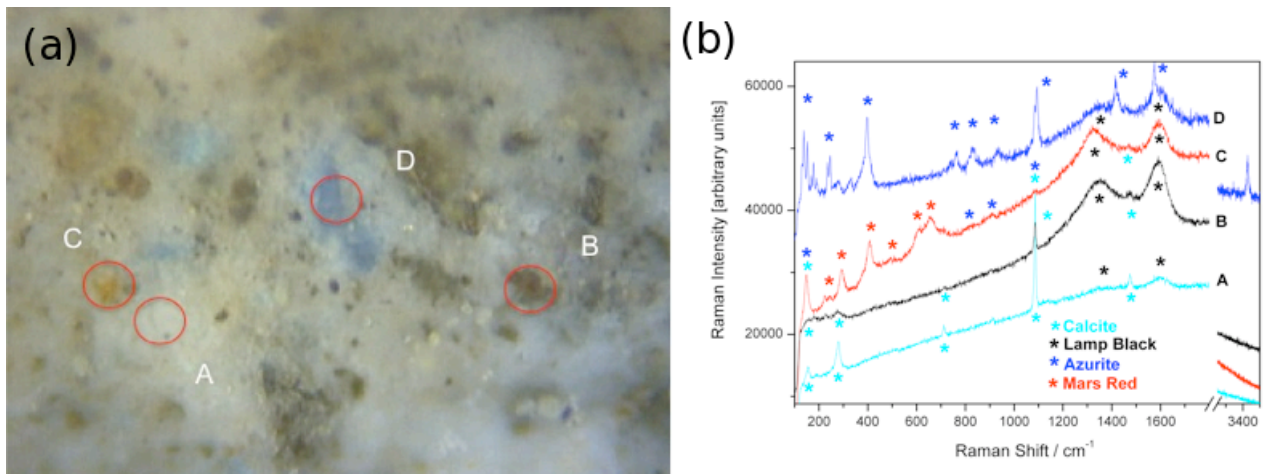

As evident from the image the different layers components were dispersed and pressed and it gave the possibility to select points that differed in color: A where white was prevalent, B representative of the dark area, C of the orange one and D for the blue zone. The laser beam was then shined over these selected spots to analyze them. In Figure 4 (b) are presented the results obtained on the selectively excited zones presented in Figure 4 (a): the spectra this time are rich in peaks and the Raman identification resulted possible, taking also advantage of a recognizable color from the microscopic image. Spectrum A (Figure 4 (b)), relative to the white spot A (Figure 4 (a)) allowed to identify $CaCO_3$ in calcite phase; calcite bands are marked with asterisks of the same spectrum cyan color. Other two bands as contaminants, marked with black asterisks come from carbon (prevalent spectrum B- dark spot B), called in the practice Carbon Black or Lamp Black [1- 4]. Following the same philosophy Red spectrum C-orange spot C was prevalent in signals from iron oxide $Fe_2O_3$ historically called Mars Red or Mars Orange [4] and being part of the Red Ochre natural pigments [1]. D spectrum-blue spot D was representative, (also here with some "contaminations" from other components and some additional bands in the low wavenumber region that may come from impurities) of the natural blue pigment $2CuCO_3Cu(OH)_2$ called Azurite, a copper based compound. The different spectra corresponding to different pigments were quite selective towards a single component and allowed the comparison with reference pure materials spectrum from Raman libraries [1- 4] (TS1): this sampling technique in spite being more destructive and loosing the cross sectional character resulted successful.

## 3. Discussion

After the microscopic inspection it was possible to formulate some preliminary hypothesis. Among the blue inorganic pigments [5, 6] are Lapislazuli and Azurite: the first, very expensive in the Medieval age was often substituted by the more economic, but still expensive, Azurite. Lapislazuli from the natural mineral Lazurite (called also Ultramarine Blue, Armenian Blue) is a sodium and aluminum silicate compound containing sulphides, calcareous spar and pyrite traces. Its approximated composition is $3Na_2O \cdot 3Al_2O_3 \cdot 6SiO_2 \cdot 2Na_2S$. Azurite, a copper basic carbonate of natural mineral origin, $2CuCO_3 \cdot Cu(OH)_2$ was the most important and used blue pigment in the old age up to XVII century. Another antique important inorganic pigment is the Egyptian Blue, approximately $CaO \cdot CuO \cdot 4SiO_2$, a copper and calcium silicates mixture, of artificial origin. Its color is very similar to that of Azurite. It was widely used in all the old age, especially in the Egyptian period. It is thus important to discriminate about the blue pigments for knowledge of the actual painting layer composition. From the microscopic visual inspection is therefore crucial to identify the blue pigment with a more specific technique especially in the present case, where no information about the artwork, for example it placement in time and space was present. Raman microscopy on surface virgin sample failed and the fragmentation of the sample in an attempt to explore an inner stratum was mandatory (Figure 4 (a)), since it was evident (Figure 2 (a) and (b)) that a blue pigment was present and its identification instead resulted not satisfactory at the surface (Figure 2 (b)). Spectra acquired selectively on the blue zone allowed obtaining enough spectral information for the Raman identification (Figure 4 (b) spectrum D) and by comparison with blue pigments spectra in published Raman libraries [1- 4] (TS1) Azurite was finally recognized.

From the cross section image in Figure 1 (b) it was shown that under the more external blue layer a orange and dark zone, and a more profound white-yellowish stratum were present and KBr

Raman spectra in Figure 4 (b) allowed to identify Carbon Black and iron oxide $Fe_2O_3$ –based pigment, of orange-yellowish color. The latter iron oxide component, from Raman libraries [1- 4] may be called Mars orange, or Mars red and comes into the composition of natural Red Ochre ($Fe_2O_3$, clay and silica) [1]. The back tint, here we called Carbon Black usually may be commonly ascribed also to Lamp Black: basically it is constituted of carbon and it is difficult to distinguish between them from the reference spectra. Another black based on carbon is Ivory Black with a similar Raman spectrum, but in this case the phosphate Raman band presence is typical and permits discrimination. Very recently in the literature appeared a study in which discrimination among the different carbon-based black was performed carefully examining in shape the two characteristic Raman bands around 1347 and 1590 $cm^{-1}$ with statistical correlation methods [8].

From the literature, giving historical information also on the use in combination of pigments [5, 6] it was possible to find out the practical reason of such preparatory layer of Azurite: it allowed the use of Azurite on fresco, which otherwise would not have been suitable for Azurite use, because the substrate would chemically attack it [5]. The common name of this Carbon Black- $Fe_2O_3$ based pigment, used as preparatory layer, is "Morellone" [5, 6]. It permitted to avoid the use of the more expensive Lapislazuli for the blue tint. Therefore it can be safely supposed that the unknown origin artwork fragment came from a fresco [5]. Usually this information about the origin or typology of an artwork is known from the beginning, when the sampling for successive analysis is performed.

The back crust is a sign of deterioration, and it can come from contaminations from the environment, of inorganic, organic or/and biological origin, more than coming of Carbon Black since during the Raman scanning of the virgin surface (Figure 2 (a), (b)) it was not possible to identify the two broad bands (around 1347 and 1590 $cm^{-1}$) evidenced instead in the Raman spectrum B of Figure 4 (b).

The identification of the deteriorated surface treatment is of interest for conservators and restorers too, since it can give indication on the best removal strategies to be adopted [5, 6]. In the case that a specific identification is not performed usually restorers have a practical experience of different removal techniques [5, 6]. With Raman microscopy, as it can be seen in Figure 2 (a) the Raman spectra in (b) it could not be straightforwardly identified with sufficient accuracy any specific surface component: most suitable is in such cases the use of infrared (IR) microscopy [9] which is more sensible towards organic components and is not disturbed by fluorescence or more in general luminescence, which in Raman spectra may be an indication of the presence of an organic deteriorated element. A luminescent background signal is also identified in Raman spectra of Figure 4 (b), since the external stratum was fragmented and included in the KBr matrix, so it was still present, moreover it may come also from binders.

## 4. Conclusion

In the present study, from the use of the Raman Microscopy technique, coupled with microscopic imaging it was possible to identify the nature and the composition of the pictorial stratum of an unknown work of art. A careful use of the method trough the sample manipulation permitted to perform a sensible analysis. This study would encourage the use of the technique and explores its capability even when its straightforward application would give not interpretable results. Raman microscopy, through its imaging capability aids to select areas of different color to obtain some separation of components in a mixed painted stratum. Its use is quite intuitive and can have enough sensitivity towards pigment recognition, while it was demonstrated to be quite weak in identifying components such deteriorated organics treatments of the surface. It was moreover evidenced how much surface treatments can give out a failure of this powerful identification system.

## 5. Materials and methods

*5.1. KBr pellet preparation*

The sample was crushed in a mortar and mixed with KBr then pressed to form a thin pellet of about less of a cm in diameter and less than a mm in thickness. To prepare the pellet a dedicated press was employed.

*5.2. Resin embedded fragment preparation*

A portion of the sample was embedded in a polyester resin as described in reference [7].

*5.3. Images acquisitions*

All the images presented in the section were acquired with the Leica microscope coupled to the Raman spectrometer. A 50X objective was employed, except in Figure 1 (a) where a 10X magnification was used. The microscope is equipped with a camera for imaging and images acquisition. It was possible to image also the laser spot impinging the sample that here, under a 50X magnification has a diameter of about 20 microns (Figure 2 (a) and 3 (a)) to have an idea of the spatial resolution in comparison with typical sizes of pigment fragments in the sample, see Figure 4 (a), where the indicated circles around the different areas has about the laser spot diameter.

*5.4. Raman spectra acquisition*

The Raman spectra (together with the images) where acquired with a commercial Renishaw Invia Raman Microscopy system working in backscattering configuration. The laser excitation came from a He-Ne laser at the 633 nm red emission. The scattered light is edge filtered and is analyzed with a single monochromator with a grating of 1800 grooves/mm spectrometer equipped with a Peltier cooled CCD camera. The microscope objective used for the selective excitation was 50X, giving a laser spot diameter of about 20 microns. The spectral resolution was 4 $cm^{-1}$, with 100 microns slit aperture. The power on the sample was ranging from 4 to 8 mW, checked not modifying the sample and being suitable for a good signal to noise ratio. All the spectra were collected at room temperature. Spectral acquisition time was 10 seconds and spectra were mediated over 3 measurements. Spectra were simply rescaled without background subtraction. Samples were placed under the microscope on a glass slide.

## 6. Supplementary materials

A table, TS1: Raman wavenumbers from the present work and from the consulted Raman libraries are reported.

**Acknowledgments**

The author acknowledge Dr. Favaro M. and Dr. Vigato P. A. (ICIS-CNR) for the virgin sample and for the resin embedded sample, Prof. Di Noto V. (University of Padova) for the preparation of the KBr pellet.

**Conflicts of Interest**

The author declare no conflict of interest

**References and Notes**

1. Bell, I. M.; Clark R. J. H.; Gibbs P. J. Raman spectroscopic library of natural and synthetic pigments (pre- ~ 1850 AD). *Spectrochimica Acta A* **1997**, 53, 2159-[2179.

2. Degen, I. A.; Newman, G. A. Raman spectra of inorganic ions. *Spectrochimica Acta A* **1993**, 49A, 859-887.

3. Bouchard et al. 2003 Bouchard, M.; Smith, D. C. Catalogue of 45 reference Raman spectra of minerals concerning research in art history or archaeology, especially on corroded metals and coloured glass. *Spectrochimica Acta A* **2003**, 59, 2247-2266.

4. Burgio, L.; Clark, R. J. H. Library of FT- Raman spectra of pigments, minerals, pigment media and varnishes, and supplement to existing library of Raman spectra of pigments with visible light excitation. *Spectrochimica Acta A* **2001**, 57, 1491-1521.

5. Campanella, L.; Casoli, A.; Colombini, M. P.; Marini Bettolo, R.; Matteini, M.; Migneco, L. M.; Montenero, A.; Nodari, L.; Piccioli, C.; Plossi Zappalà, M.; Portalone, G.; Russo, U.; Sammartino, M. P. *Chimica per l'arte*, Zanichelli Ed. Italy, 2007.

6. Matteini, M.; Moles, A. *La chimica nel Restauro-i materiali dell'arte pittorica*. Nardini Ed. Florence, Italy, 1989.

7. Gambirasi, A.; Peruzzo, L.; Bianchin, S.; Favaro, M. Electron backscatterer diffraction in conservation science: phase identification of pigments in paint layers. *Microsc. Microanal.* **2013** 19, 921-928.

8. Tomasini, E. P.; Halac, E. B.; Reinoso, M.; Di Liscia, E. J.; Maiera, M. S. Micro-Raman spectroscopy of carbon-based black pigments. *J. Raman Spectrosc.* **2012**, 43, 1671–1675.

9. Favaro, M.; Bianchin, S.; Vigato, P. A.; Vervat, M. The palette of the Macchia Italian artist Giovanni Fattori in the second half of the XIX century. *J. Cultural Heritage* **2010**, 11, 265-278.